\documentclass[preprint,5p,times,twocolumn]{elsarticle}
\usepackage{graphicx}
\usepackage{amsmath,amssymb}
\usepackage{subfig} 
\usepackage{xcolor}
\usepackage{chemformula}
\usepackage{ulem}
\usepackage{physics}
\usepackage{mhchem}

\newcommand{\ba}{\begin{eqnarray}}
\newcommand{\ea}{\end{eqnarray}}
\newcommand{\be}{\begin{equation}}
\newcommand{\ee}{\end{equation}}

\newcommand\redsout{\bgroup\markoverwith{\textcolor{red}{\rule[0.5ex]{2pt}{0.4pt}}}\ULon}

\begin{document}

\title{The presence of non-analyticities and singularities in the wavefunction and the role of "invisible" delta potentials}

	\author[1]{Jorge Munzenmayer}  
	\ead{jorge.munzenmayer@alumnos.usm.cl}
	
    	\author[2]{Derek Frydel} 
    	\ead{derek.frydel@usm.cl}

    	\address{Department of Chemistry, Federico Santa Maria Technical University, Campus San Joaquin, 7820275, Santiago,Chile}

\begin{abstract}
This article examines the suggestion made in Ref. \cite{Huang16} that a solution to a particle in 
an infinite spherical well model, if it is square-integrable, is a physically valid solution, even if at 
the precise location of the singularity there is no underlying physical
cause, therefore, the divergence would have to be a nonlocal phenomenon caused by 
confining walls at a distance.  In this work we examine this claim more carefully.  
By identifying the correct differential equation for a divergent square-integrable solution and 
rewriting it in the form of the Schr\"odinger equation, we infer that the divergent wavefunction 
would be caused by the potential $V(r) \approx - r\delta({\bf r})$, which is a kind of  
attractive delta potential.  Because of its peculiar form and the fact that 
it leads to a divergent potential energy $\langle V\rangle=-\infty$, the potential $V(r)$ and the divergent 
wavefunction associated with it are not physically meaningful.  
\end{abstract}

\maketitle

\section{Introduction}

The present article 
starts out as an examination of a 
claim made in Ref. \cite{Huang16} that a divergent wavefunction represents a valid physical 
state for a particle in a spherical well model, a model that apart from confining walls has no other 
potential that could be regarded as a physical cause of a singularity.  This essentially would 
make a singularity a non-local phenomenon.  In addition to directly engaging with the claim
in Ref. \cite{Huang16}, the article examines other "physically valid" cases where some sort
of non-analyticity arises and looks carefully into mathematical structure associated with that 
non-analyticity, as well as into a general question:  is a divergent wavefunction physically 
meaningful in any system?


Besides mathematical concerns, there are physical consequences of accepting a divergent 
solution as a physical wavefunction.  For example, 
by changing the shape of the confining well from a cube to a sphere we would give rise to 
a state with a divergent wavefunction -- since we know of no divergent wavefunction for a particle in 
a box model.  

Besides quantum mechanics, there are many other 
physical systems described by the wave equation.  For example, standing acoustic 
waves in a spherical cavities \cite{Russella10}, or 
the standing electromagnetic waves inside a spherical cavity 
that give rise to Casimir forces.   The conclusions of Ref. \cite{Huang16} 
if correct would have far reaching repercussion.  



To summarize the claim of Ref. \cite{Huang16}, the authors indicate that a particle in a 
spherical well model, in addition to the usual class of regular solutions, represented by 
the spherical Bessel functions $j_l(kr)$, admits another class of solutions that diverge at 
$r=0$, corresponding to the spherical Neumann functions denoted as $n_l(kr)$.  

Conventionally, the functions $n_l(kr)$ are deemed unphysical on account of their divergence 
at the center of the spherical well, $n_l(r) \approx r^{-(l+1)}$, where $l=0,1,\dots$.  The authors
note, however, that since the relevant physical quantity is the density, 
$\rho_l(r) \approx  r^{-2(l+1)}$, the correct criterion should not be whether or not a 
wavefunction diverges, but rather that it be 
square-integrable.  Such a divergent but square-integrable solution happens to be for $l=0$, 
in which case the Neumann function becomes $n_0(kr)=-\cos(kr)/(kr)$.  The authors then 
include this solution among valid states of the system.  
And as square-integrability is the sole criterion of physicality, no concern is given about  
a physical cause underlying the occurrence of the singularity and/or a mathematical 
term that gives rise to it.  
 





This paper is organized as follows.  We start with the discussion in Sec. (\ref{sec:cusp}) 
of the Coulomb cusp condition, as a well known example of a system with a wavefunction 
containing non-analyticity.   In Sec. (\ref{sec:cusp2}) we look into the cusp from a different 
angle, by reformulating the Schr\"odinger equation and introducing an "invisible" delta
potential that strictly speaking is correct but unnecessary.  Yet maintaining it in the 
formalism leads to rigorous link between the hydrogen atom and the delta-potential model.  
In Sec. (\ref{sec:box3d}) we consider the divergent solution for the particle in
an infinite spherical well model and identify the correct differential equation associated
with that solution.  By rewriting this equation in the form of the Schr\"odinger equation, 
we infer the delta potential at the location of the divergence and that gives rise to it.

\section{Coulomb cusp condition}
\label{sec:cusp}

As a starting point, we review the case of a Coulomb cusp condition where the cusp
in the wavefunction can be considered as a weak singularity (or a non-analyticity)
whose presence can be traced to the Coulomb potential and the position of the proton.   
The well defined link between the non-analyticity and its physical cause are formally 
expressed in the relation known as the Kato's cusp condition or the Kato theorem \cite{Kato57}.

The Schr\"odinger equation for the hydrogen atom model is
$$
\bigg[-\frac{\hbar^2}{2m}\nabla^2  - \frac{Ze^2}{4\pi\epsilon }\frac{1}{r} \bigg] \psi(r) = E \psi({r}).  
$$
Being interested in the ground state, the wavefunction of which is spherically symmetric, we 
consider only the radial part of the Laplacian operator, for an arbitrary dimension $D$ given by 
\cite{Muslih11,Stillinger77}
\be
\nabla^2 \to \bigg[\frac{\partial^2}{\partial r^2} + \frac{D-1}{r} \frac{\partial }{\partial r}\bigg], 
\label{eq:LD}
\ee
so that the Schr\"odinger equation simplifies to
\be
\frac{\partial^2 \psi({r})}{\partial r^2} + \frac{D-1}{r} \frac{\partial \psi({r})}{\partial r} + \frac{2Z}{a_0}\frac{1}{r}\psi(r) = k^2 \psi({r}), 
\label{eq:SED}
\ee
where 
$$
k = \sqrt{-\frac{2mE}{\hbar^2}},
$$
is a positive number (since $E<0$) and $a_0 = \frac{4\pi\epsilon \hbar^2}{me^2}$ is the Bohr radius. 
A $D$-dependent  Schr\"odinger equation might be of interest to systems under strong confinement, 
which effectively reduce dimensionality.

In the neighborhood of the point $r=0$, the Schr\"odinger equation in Eq. (\ref{eq:SED}) is completely 
dominated by the terms proportional to $1/r$, 
\be
\frac{\partial \psi({r})}{\partial r} + \frac{Z}{a_0} \frac{2}{D-1} \psi(r) = 0, ~~~~~ \text{as}~~ r\to 0.  
\label{eq:cusp0}
\ee
After rearranging the above result, we get the cusp condition as it is known in its more familiar form 
\be
\frac{1}{\psi(0)} \frac{\partial \psi({r})}{\partial r}\bigg|_{r=0} = - \frac{Z}{a_0} \frac{2}{D-1}.  
\label{eq:cusp}
\ee

It turns out that the ground state wavefunction satisfies Eq. (\ref{eq:cusp0}) not only in the 
neighborhood of $r=0$ but in the entire range of $r$.  Since the solution to Eq. (\ref{eq:cusp0}) is
\be
\psi_D(r) \propto e^{-\frac{r}{a_0}\frac{2Z}{D-1}}, 
\label{eq:psi_D}
\ee
inserting this result to Eq. (\ref{eq:SED}) yields 
$$
k = \frac{2Z}{a_0(D-1)}, 
$$
so that the ground state energy is 
\be
E = -\frac{2Z^2\hbar^2}{m a_0^2} \bigg(\frac{1}{D-1}\bigg)^2.  
\label{eq:E}
\ee 
The above expression indicates that the more degrees of freedom the system has, the higher 
the energy of its ground state.  In the limit $D\to\infty$ the electron in the ground state should 
become unbounded.  On the other hand, $E$ diverges as $D\to 1$, and judging from the 
wavefuntion in Eq. (\ref{eq:psi_D}) and the corresponding density $\rho(r) = |\psi(r)|^2$, the 
electron appears to be collapsing into proton \cite{Loudon59,Loudon16}.

In the case of $D=1$, instead of the Coulomb potential,  there is another bounding potential 
which produces an exponential wavefunction and the cusp condition. This is the delta potential 
$V(x) = -\alpha\delta(x)$ and the Schr\"odinger equation that results is 
\be
\psi''(x) + \alpha'\delta(x)\psi(x) = k^2\psi(x),~~~ \text{where}~~~ \alpha' = \frac{2m\alpha}{\hbar^2}.  
\label{eq:S1D}
\ee
The cusp condition is obtained by operating on this equation with the operator $\lim_{a\to 0}\int_{-a}^{a}dx\,$ 
at the location of the delta potential leading to  
\be
\frac{1}{\psi(0)} \frac{d \psi(x)}{dx}\bigg|_{x=0^+} = - \frac{\alpha'}{2}.   
\ee

\section{The second look at the Coulomb cusp condition}
\label{sec:cusp2}

In this section we take a little different view of the cusp condition.  Since both the hydrogen 
atom and the delta potential model produce the same type of non-analyticity, we look for 
a deeper connection between the two models.  The idea is to look for the presence of a hidden 
delta function in the Schr\"odinger equation at the location of a cusp, that from the physical point 
of view may play no role but is useful for mathematical consistency.   

Consider the ground-state wavefunction for the hydrogen atom model for a general $D$-dimension, 
first shown in Eq. (\ref{eq:psi_D}).  
For the case $D=3$ the dominant terms in the neighborhood $r=0$ are 
$$
\psi(r) \propto  1 - \bigg(\frac{Z}{a_0}\bigg)\, r + \dots.  
$$
Because the expansion does not include the term $r^{-1}$,  
it is assumed that that the Laplacian equation of that solution does not include the delta function.    

To demonstrate that strictly speaking this is not the case, we define the identity 
$$
1=\frac{r}{r},
$$
then take the Laplacian of its both sides, 
$$
0 = \nabla^2 \bigg(\frac{r}{r}\bigg).  
$$
After expanding the right hand side we get   
$$
 0 = r \nabla^2\bigg(\frac{1}{r}\bigg) +  \frac{1}{r} \nabla^2 r + 2\grad \bigg(\frac{1}{r}\bigg) \cdot \grad r, 
$$
then using the identity $\nabla^2 r^{-1} = -4\pi\delta({\bf r})$ and 
$\grad r = \hat {\bf r}$ and $\grad r^{-1} = -\hat{\bf r} r^{-2}$, we get 
\be
\nabla^2 r = \frac{2}{r} + 4\pi r^2 \delta({\bf r}), 
\ee 
As the first term can be obtained by straightforward application of 
the spherically symmetric part of the Laplacian operator, 
$\big(\frac{\partial^2}{\partial r^2} + \frac{2}{r} \frac{\partial }{\partial r}\big) r$, 
the emergence of the delta function is less obvious.  

What essentially this means is that the Laplacian of the wavefunction 
$\psi(r)\propto e^{-\frac{Zr}{a_0}}$ is
\be
\nabla^2 \psi(r) = k^2\psi - \bigg(\frac{4\pi Z}{a_0}\bigg) r^2 \delta({\bf r}), 
\label{eq:delta_H}
\ee
and by rewriting this as the Schr\"odinger equation we get 
we would write 
\be
\bigg[-\nabla^2 - \frac{2Z}{a_0}\frac{1}{r} - \frac{Z}{a_0}  4\pi r^2 \delta({\bf r}) \bigg]\psi(r) = k^2\psi(r), 
\ee
so that the potential energy term is 
$$
V(r) = - \frac{Z \hbar^2 }{2ma_0}  4\pi r^2 \delta({\bf r}),
$$
such that 
\ba
\langle V\rangle &=& -\frac{Z\hbar^2}{2m a_0} \int d{\bf r}\, 4\pi r^2 \psi^2({\bf r}) \delta({\bf r}) 
\nonumber\\ &=& -\frac{Z\hbar^2}{2m a_0} \lim_{r\to 0} 4\pi r^2 \psi^2(r) = 0.
\ea
The potential $V(r)$, therefore, is completely invisible.  

We can generalize the result in Eq. (\ref{eq:delta_H}) to any dimension.  
$$
\nabla^2 r =  \frac{D-1}{r} +  \Omega_ D r^{D-1} \delta({\bf r}). 
$$
where 
$$
\Omega_D = \frac{2\pi^{D/2}}{\Gamma(D/2)} 
$$
is the $D$-dependent solid angle factor such that $\Omega_1=2$, $\Omega_2=2\pi$, and $\Omega_3=4\pi$.  
The generalized $D$-dependent Schr\"odinger equation then becomes 
\be
\bigg[-\nabla^2 - \frac{2Z}{a_0}\frac{1}{r} - \frac{2Z}{a_0}  \frac{\Omega_D}{D-1} r^{D-1} \delta({\bf r}) \bigg]\psi_D(r) = k^2\psi_D(r), 
\label{eq:SED_delta}
\ee
so that the $D$-dependent delta potential is 
\be
V_D(r) = - \frac{Z\hbar^2 }{ma_0}  \frac{\Omega_d r^{D-1}} {(D-1)} \delta({\bf r}).  
\label{eq:VDH}
\ee




Up to this point, 
it is not clear what advantage there is to using the Schr\"odinger equation in 
Eq. (\ref{eq:SED_delta}), since the potential 
$V_{D}(r)$ in Eq. (\ref{eq:VDH}) is "invisible" and not a part of a physical system.  
We will next show that including the "invisible" delta potential in Eq. (\ref{eq:SED_delta})
can lead to some mathematical consistencies.  

If we scale the Coulomb potential as 
as
$$
Z' = Z(D-1),
$$
so that the Schr\"odinger equation for the hydrogen atom for an arbitrary dimension $D$ becomes 
\be
\bigg[-\nabla^2 - \frac{2Z(D-1)}{a_0}\frac{1}{r} - \frac{2Z}{a_0}  \Omega_D r^{D-1} \delta({\bf r}) \bigg]\psi(r) = k^2\psi(r), 
\ee
and then continuously change $D$ from $D=3\to 1$, even if we transform the Schr\"odinger 
equation, the wavefunction remains the same $\psi(r) \propto e^{-\frac{2Z r}{a_0}}$.  
Because for $D=1$ the Coulomb potential vanishes, it is not clear what physical feature
of the Schr\"odinger equation produces the exponential wavefunction.  It happens that
at exactly $D=1$ the delta potential in Eq. (\ref{eq:VDH}) becomes "visible",  
\be
\bigg[-\frac{d^2}{dx^2} - \frac{4Z}{a_0} \delta(x) \bigg]\psi(x) = k^2\psi(x),
\label{eq:SD1a}
\ee
and because the physical cause of the exponential wavefunction and the cusp condition.


The delta potential system in Eq. (\ref{eq:SD1a}) is frequently referred to as the hydrogen 
atom, because the wavefunction of its bound state has the same functional form 
as the ground-state wavefunction of the hydrogen atom in $D=3$.  The mathematical 
connection established above, through the combined presence of an "invisible" delta 
and the Coulomb potential and the application of dimensional transformation 
reveals a deeper mathematical link.  



\section{Particle in a spherical box model}
\label{sec:box3d}

We next turn to a particle in an infinite spherical well model governed by the following 
Schr\"odinger equation 
\be
\nabla^2 \psi({\bf r}) = -k^2 \psi({\bf r}), ~~~~ \text{~for~~~} |{\bf r}|\le a,
\label{eq:S1}
\ee
where 
$$
k =  \sqrt{\frac{2mE}{\hbar^2}}, 
$$
and $E>0$.  Anywhere outside the spherical well, $r\ge a$, the wavefunction 
vanishes.  

Considering only spherically symmetric solutions which include the ground state, the 
Sch\"ordinger equation for $D=3$ reduces to 
\be
 \frac{\partial^2 \psi({r})}{\partial r^2} + \frac{2}{r} \frac{\partial \psi({r})}{\partial r}  = -k^2 \psi({r}).  
\label{eq:SED2}
\ee
The two families of possible spherically symmetric solutions, as pointed out in Ref. \cite{Huang16}, 
are
\be
\psi_R(r) \propto \frac{\sin(k_nr)}{k_nr}, \text{~~~~}  k_n = \frac{n\pi}{a},  \text{~for~~~} n=1,2,\dots,
\label{eq:solA}
\ee
and 
\be
\psi_S(r) \propto \frac{\cos(k_nr)}{k_nr},  \text{~~~~}  k_n = \frac{(2n-1)\pi}{2a},  
\text{~for~~~} n=1,2,\dots,
\label{eq:solB}
\ee
where the ground state corresponds to the principal number $n=1$.   Even though both 
solutions vanish at $r=a$, they have very different properties in the neighborhood $r=0$.  
The solution $\psi_R$ is regular in the neighborhood $r=0$, 
$$
\psi_R(r) \propto\frac{\sin(kr)}{kr} = 1- \frac{k^2r^2}{6} + \dots, 
$$
while the solution $\psi_S$ in the same neighborhood has singularity
$$
\psi_S(r) \propto\frac{\cos(kr)}{kr} = \frac{1}{kr} - \frac{kr}{2} + \dots.  
$$

Both solutions $\psi_R$ and $\psi_S$ can be constructed from the functions $\frac{e^{\pm ik r}}{kr}$, 
whose Laplacian is given by 
\be
\nabla^2  \bigg(\frac{e^{\pm ikr}}{k r}\bigg) = -k^2 \bigg(\frac{e^{\pm ikr}}{k r}\bigg) -  \frac{4\pi}{k} \delta({\bf r}), 
\label{eq:L3}
\ee
and where the delta function is the result of the identity 
\be
\nabla^2 \bigg(\frac{1}{4\pi r}\bigg) = -\delta({\bf r}).  
\label{eq:delta}
\ee
In the case of the regular solution $\psi_R$, the delta function is cancelled out, 
\be
\nabla^2  \bigg(\frac{\sin k r}{k r}\bigg) = -k^2 \bigg(\frac{\sin k r}{k r}\bigg), 
\ee
meaning that $\psi_R$ is a true solution to Eq. (\ref{eq:S1}).  On the other hand, the solution 
$\psi_S$ retains the delta function, 
\be
\nabla^2  \bigg(\frac{\cos k r}{k r}\bigg) = -k^2 \bigg(\frac{\cos k r}{k r}\bigg)  -  \frac{4\pi}{k} \delta({\bf r}), 
\label{eq:diff2}
\ee
and, in consequence, $\psi_S$ cannot be considered solution of the original system 
in Eq. (\ref{eq:S1}).  It is rather a solution of an alternative differential equation given by 
\be
\nabla^2 \psi_S({\bf r})  =  - k^2 \psi_S({\bf r}) - \frac{4\pi}{k} \delta({\bf r}).  
\label{eq:S3a}
\ee

A related differential equation is encountered in other physical situations.  Within 
the Debye-H\"uckel theory of electrostatics, the electrostatic potential $\phi$ is governed 
by the following linear differential equation \cite{Yan02,Frydel16}
\be
\nabla^2 \phi({\bf r})  =  \kappa^2 \phi({\bf r}) - 4\pi\lambda_B \delta({\bf r}),
\label{eq:DH}
\ee
where $\kappa$ is the screening parameter and $\lambda_B$ is the Bjerrum length.  The delta
function in the above equation has a definite physical interpretation and represents a point 
charge located at the origin.  The solution is the screened electrostatic 
potential at the location of a point charge, $\phi(r) = \frac{\lambda_B e^{-\kappa r}}{r}$.  
If $\psi_S$ were to be taken as a physically relevant solution, then the delta 
function in Eq. (\ref{eq:S3a}) would need to have a clear physically motivated cause.

To arrive at a physical interpretation of the delta function in Eq. (\ref{eq:S3a}), we rewrite Eq. (\ref{eq:S3a})
so that it takes on the form of the Sch\"ordinger equation,  
\be
\bigg[-\nabla^2 - \frac{4\pi}{k} \frac{\delta({\bf r})}{\psi_S(0)} \bigg] \psi_S({\bf r}) = k^2 \psi_S({\bf r}).  
\label{eq:S1b}
\ee
Since in the neighborhood $r=0$, $\psi_S(r) \propto (kr)^{-1}$, the above equation can be 
further modified into 
\be
\Big[-\nabla^2 - 4\pi r \, \delta({\bf r}) \Big] \, \psi_S({\bf r}) = k^2 \psi_S({\bf r}).  
\label{eq:S1c}
\ee
The potential energy that emerges, 
\be
V({\bf r}) =   -\frac{\hbar^2}{2m} 4\pi r\delta({\bf r}), 
\label{eq:V}
\ee
is a kind of an attractive delta potential, apart for the factor $r$.  At first sight the potential 
appears "invisible", since at the origin $r$ vanishes and should have no effect
on observable quantities.  However, since any physical quantity depends on 
the density $\rho(r)=|\psi_S(r)|^2$, and since this density diverges as 
$r^{-2}$, the average value of the potential is not zero but quite contrary, it diverges 
\ba
\langle V\rangle &=& -\frac{2\pi \hbar^2}{m} \int d{\bf r}\, r \psi_S^2({\bf r}) \delta({\bf r}) 
\nonumber\\ &=& 
-\frac{2\pi \hbar^2}{m} \lim_{r\to 0} r \psi_S^2({\bf r}) 
=-\frac{2\pi \hbar^2}{m k_n^2} \lim_{r\to 0} \frac{C_n^2}{r},
\ea
where $C_n$ is the normalization factor, $\psi_S = C_n \cos k_n r / k_n r$, given by 
$$
C_n = (2n-1)\sqrt{\frac{\pi}{8a^3}}.  
$$
Despite the divergent potential energy, the total energy is finite, indicating that the kinetic 
energy has its own positive divergence that exactly cancels out the negative divergence 
of the potential energy.  The total energy is given by 
$$
E_n =  \frac{k_n^2 \hbar^2}{2m},
$$
where $k_n$ is defined in Eq. (\ref{eq:solB}).

If we compare the ground state energies corresponding to the principal number $n=1$ for both 
wavefunctions $\psi_R$ and $\psi_S$ and their corresponding Hamiltonians we get
$$
E_1^R =  \frac{\pi^2 \hbar^2}{2ma^2}, ~~~~~ E_1^S =  \frac{\pi^2 \hbar^2}{8ma^2},
$$
so that 
$$
E_1^R = 4 E_1^S.  
$$
The fact that $E^S<E^R$ can be traced to the presence of the attractive potential $V(r)$, 
which lowers an overall energy but fails to trap a particle, even if it produces a divergent 
wavefunction.  The average position of a particle in each system is 
$$
\langle r\rangle_R = \frac{1}{2}, ~~~~~ \langle r\rangle_S = \frac{1}{2} - \frac{2}{\pi^2},
$$
so that the average position for the system with the potential $V(r)$ occupies a more
central position.

It is difficult to imagine that the potential $V(r)$ in Eq. (\ref{eq:V}) represents an actual 
physical system, (1) due to its peculiar functional form, in particular, the fact that it vanishes 
at the origin, (2) because it leads to a divergent potential energy $\langle V\rangle$, (3) and it 
has no independent parameter of strength, namely, its magnitude is completely determined 
by the particle mass.  This lack of flexibility does not make $V(r)$ a good candidate for a 
physical potential and implies that it is a mathematical oddity that arises by imposing the 
physical meaning to the divergent wavefunction $\psi_S(r)$.  

\section{Hydrogen atom in an infinite spherical well:  unbounded states}

As the last example of an example where a divergent solution may arise, 
we consider a hydrogen atom trapped in an infinite spherical well.  
A hydrogen molecule in confinement has been of interest 
since early days of quantum mechanics \cite{Boer37,Sommerfeld38,Groot46}, 
and in recent years it received renewed interest due to its connection 
to realistic systems, one prominent example being quantum dots
\cite{Weil76,Ludena77,Cooper00,Hall08}.  
Like the particle in a spherical well, this system, too, admits of a divergent solution
so that one could attempt to force physical interpretation.

The maximum value of $Z$ at which a particle remains unbounded by the Coulomb potential 
occurs when the negative potential energy is completely cancelled 
by the kinetic energy,  $\langle V\rangle = -\langle K\rangle$.  This means $E=0$, or 
$k=0$, in which case the Schr\"odinger equation in Eq. (\ref{eq:SED}) for the case $D=3$ reduces to 
\be
\frac{\partial^2 \psi({r})}{\partial r^2} + \frac{2}{r} \frac{\partial \psi({r})}{\partial r} + \frac{2Z}{a_0}\frac{1}{r}\psi(r) = 0.  
\label{eq:SEDW}
\ee
The non-divergent ground-state solution is 
\be
\psi(r) \propto   
\frac{j_1\big( \sqrt{8Zr/a_0} \big)}{\sqrt{r}},
\label{eq:psi_RH}
\ee
where $j_m(x)$ is the Bessel function of the first kind.
The requirement that the wavefunction vanishes at $r=a$ provides us with 
the relation
$$
v_k = \sqrt{\frac{8Za}{a_0}}, 
$$
where $v_k$ are zeros of the Bessel function and the lowest zero $v_1=3.83171$ corresponds 
to the ground-state.  


The wavefunction in Eq. (\ref{eq:psi_RH}) satisfies the cusp condition in Eq. (\ref{eq:cusp}).  
This can be shown by expanding the wavefunction around $r=0$, where the initial terms are 
$$
\frac{j_1\big( \sqrt{8Zr/a_0} \big)}{\sqrt{r}} \propto 1 + \frac{Z}{a_0} r + \dots, 
$$
and which satisfy the cusp relation 
$$
\frac{1}{\psi(0)} \frac{\partial \psi({r})}{\partial r}\bigg|_{r=0} = \frac{Z}{a_0}. 
$$


A divergent solution to Eq. (\ref{eq:SEDW}) is given by 
$$
\psi(r) \propto   -\frac{y_1\big( \sqrt{8Zr/a_0} \big)}{\sqrt{r}},
$$
where $y_m(x)$ is the Bessel function of the second kind, 
and which in the region $r=0$ is dominated by the terms 
$$
-\frac{y_1\big( \sqrt{8Zr/a_0} \big)}{\sqrt{r}} \propto \frac{1}{r}  - \frac{2Z}{a_0} \ln r + \dots. 
$$
The first observation is that the divergent solution fails to satisfy the cusp condition.  
The reason is that the divergent function 
is not a solution of the original physical problem, but corresponds to the system
whose potential is
$$
V_{tot}(r) = -\frac{Z}{4\pi \epsilon r}  - \frac{\hbar^2}{2m} 4\pi r\delta({\bf r}).  
$$
As previously, the divergent solution must be rejected on physical grounds.  



There is an addition problem with the divergent solution in this case.  
The divergent solution in this case yields a divergent total energy, $E_1= -\infty$.
Considering just the Coulomb energy, 
we find that its average values diverges,
$$
-\bigg\langle \bigg|\frac{Z}{4\pi \epsilon r} \bigg|\bigg\rangle = -\frac{Z}{\epsilon}\int_0^a dr\, r |\psi(r)|^2 = -\infty.  
$$
Another divergence comes from the delta potential 
$V(r) \approx - r\delta({\bf r})$.  The kinetic energy cancels the divergence due to the 
delta potential but not that due to the Coulomb interactions and, as a consequence, the 
total energy diverges.

\section{Conclusion}
\label{sec:conclusion}



We conclude that a divergent wavefunction, even if it is square-integrable, is not a solution 
of the particle in a box model.  This means that the property of square-integrability is insufficient 
condition of physical eligibility.  Instead, the physical aspects of the system must be taken into 
account.  The divergence requires some physical cause.  We identify this cause 
as the delta like attractive potential $V(r)\approx -r\delta({\bf r})$.  This potential, however, 
lends little physical interpretation and is best regarded as a mathematical curiosity;  
first, because of its unusual functional form; and second, because it yields the average potential 
energy that diverges.  These observations lead to the conclusion that the presence of a 
divergence in a wavefunction, although an interesting feature, is highly unlikely from 
physical point of view.




\end{document}